\documentclass[sigconf]{acmart}

\usepackage{booktabs} 
\usepackage{amsmath}

\setcopyright{rightsretained}
\renewcommand\footnotetextcopyrightpermission[1]{} 
\acmYear{2018}
\copyrightyear{2018}
\begin{document}
\title{A Design Space Exploration (DSE) on Non-Invasive Sensing of Bladder Filling Using Near Infrared Spectroscopy (NIRS)}

\author{M. Saffarpour}
\affiliation{}
\email{msaff@ucdavis.edu}

\author{S. Ghiasi}
\affiliation{}
\email{ghiasi@ucdavis.edu}

\begin{abstract}
Urinary Incontinence (UI) is a widespread medical condition that affects one person from every three or four Americans. Near-Infrared Spectroscopy (NIRS) is a non-invasive under-study method for bladder filling sensation that can enhance the life quality of UI patients by finding the optimal voiding time. However, the application of NIRS to bladder volume sensing can be quite challenging due to three major obstacles: non-adequate traversal depth of NIR wavelengths, robustness and power efficiency requirements of the application, and low power transmission rate of NIR wavelengths. This work provides a Design Space Exploration (DSE) through the effect of various design parameters on NIRS applicability for bladder volume sensing. We investigate the impact of 7 different wavelengths from 650-950 nm, 16 possible detector-source distances, and 6 different sensation depths. The results of our work can be used as a guideline through optimal design and implementation of NIRS for bladder filling sensation.
\end{abstract}

%
%

\maketitle

\section{Introduction}
NIRS can be a promising method for non-invasive bladder monitoring of patients with UI symptoms. The loss of bladder control, known as UI, is a world-wide prevalent medical condition that affects 25-33\% of Americans \cite{2018:Urology}. NIRS technique is projected to be a useful tool for sensing the bladder filling which can help in finding the optimal voiding time in short-term, while providing valuable information for the long-term treatment process.

Although NIRS technique has been tested for bladder volume sensing in some clinical trials \cite{2005:macnab}\cite{2014:macnab}\cite{2014:molavi}, its applicability for practical usage faces several challenges.

First of all, we show that due to high absorption and scattering characteristics of tissue layers in abdominal area for NIR wavelengths, the traversal depth may not be adequate to sense the bladder. This problem becomes more challenging when bladder depth is higher as a result of patient's obesity.

Secondly, robustness and power efficiency of the final probe is a requisite. This device should not be sensitive to misplacements and small movements during usage and should be low power in order to be powered by battery in a daily basis.

On the other hand, the intensity of detected photons should be high enough to provide a reasonable power range to the detection module. Since the input power is limited by system's power budget, providing the minimum sensational power for detection module can be an obstacle. Therefore, the optimal wavelength should be able to maximize the power transmission ratio from input to output, and as a result, minimize the absorption and scattering power losses in the transmission process.

These challenges necessitate the need for a comprehensive design space exploration (DSE) to evaluate the effect of parameters such as wavelength and source-detector distances on penetration depth, sensitivity, and power transmission ratio. In this work, we employed monte carlo simulation to perform the aforementioned DSE while considering following parameters:
\begin{itemize}
\item 7 wavelengths in range of 650-950 nm
\item 16 possible detector-source spacings
\item 6 different tissue thicknesses in range of 15-40 mm
\end{itemize}

In what follows, first the power transmission ratio calculation has been explained (section 2). Then, section 3 and 4 cover the simulation setup and the results of this project, consecutively. Finally, Section 5 is dedicated to conclusion of this work.

\section{Power Transmission Ratio}
The power transmission ratio is the ratio of detected photons power to the input photons' power. In this work, we assume that the detection of all photons happen at the same time. As a result, the power transmission ratio would be equal to energy transmission ratio.
The energy of $N$ photons with wavelength $\lambda$ can be calculated by equation \ref{eq:equation1} where $h$ and $c$ are the Plank constant and speed of light, consecutively.
\begin{align}
E= N * \dfrac{h c}{\lambda} \label{eq:equation1}
\end{align} 
Therefore, the overall energy transmission ratio would be equal to the number of detected photons to the number of input photons.

\section{Simulation Setup}
We have used Monte Carlo Extreme simulator in order to quantify the effect of wavelength and source-detector spacing on traversal depth.

\begin{figure}\centering
	\includegraphics[width=8cm]{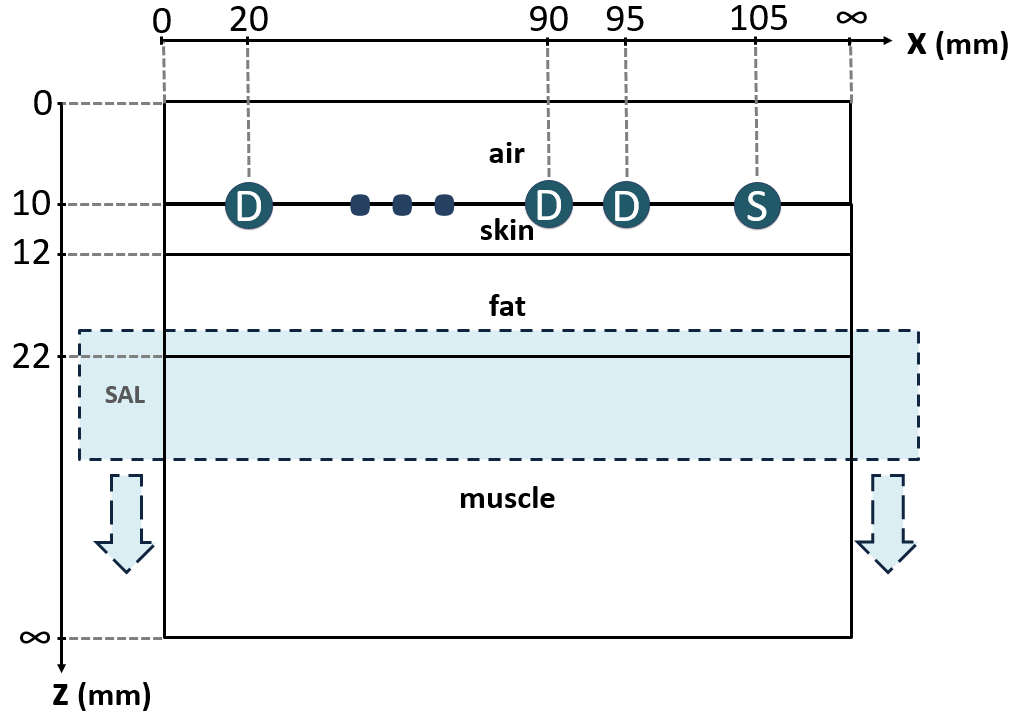}
	\caption{The simulation model which is capable of performing penetration depth analysis by using a sweeping SAL.}
	\label{fig:penetration}
\end{figure}

In order to get track of photons' traversal depth in the simulation, we have compared the number of detected photons with and without a super-absorbent layer (SAL) at the depth of interest. SAL has significantly high absorption and zero scattering characteristic which would swallow and absorb photons reaching its surface. The decreased number of detected photons, as a result of applying SAL, indicates the number of photons reaching the SAL depth. The simulation model has been presented in figure \ref{fig:penetration}. 

We utilized Monte Carlo simulation with 500 million input photons at 7 different NIR wavelengths in range of 650nm to 950nm. The optical properties of tissue layers used in this model has been adopted from \cite{1998:Simpson}.

For all the wavelengths under study, we have swept the SAL from 10 mm to 40 mm with a step-size of 5 mm and collected the penetration depth information. The simulation contains 16 detectors which are placed every 5 mm in range of 10-85 mm  distance to light source. All these 16 detectors have the similar radius of 1.41 mm.

Finally, the power transmission ratio has been calculated per wavelength for each source-detector distance using equation \ref{eq:equation1} and number of detected photons at each detector. Then we back-calculated the minimum input power required for receiving the minimum sensible output power (detector characteristics provided in \cite{2016:fairchild}) considering these transmission ratio values.

\section{Results and Discussion}
\begin{figure*}\centering
	\includegraphics[width=\textwidth]{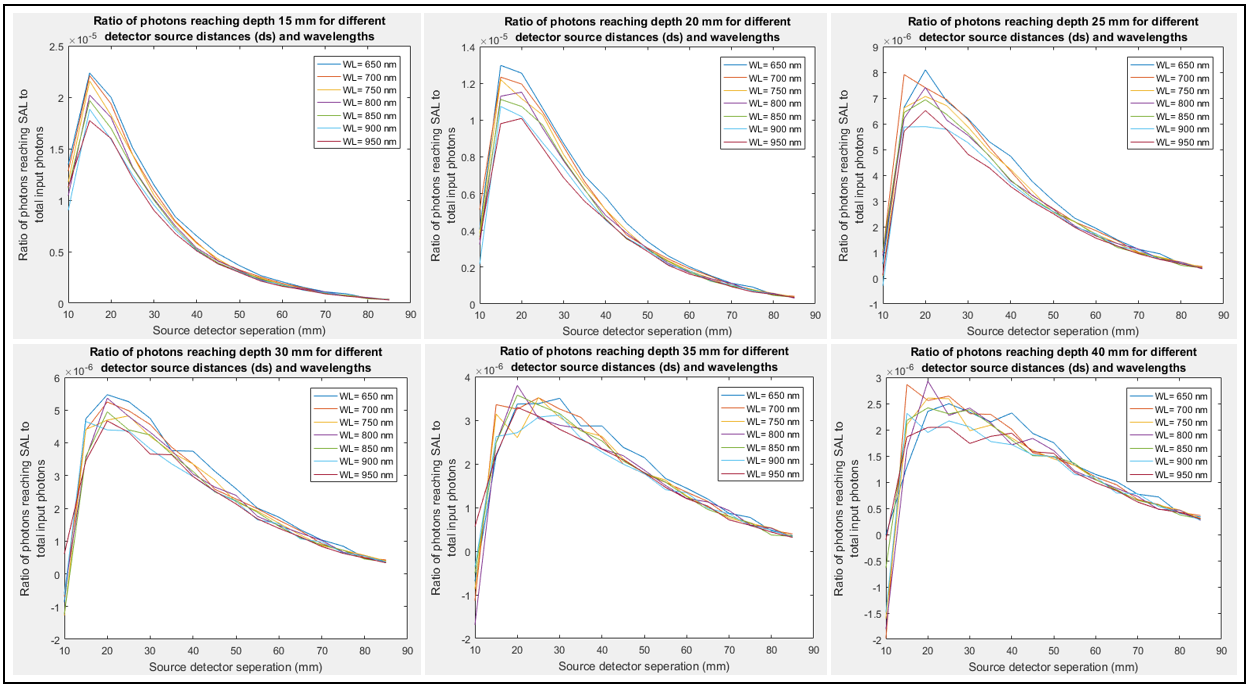}
	\caption{ The ratio of photons reaching depth of interest to overall number of input photons at each source-detector distance.}
	\label{fig:photons_reaching}
\end{figure*}

\begin{figure*}\centering
	\includegraphics[width=\textwidth]{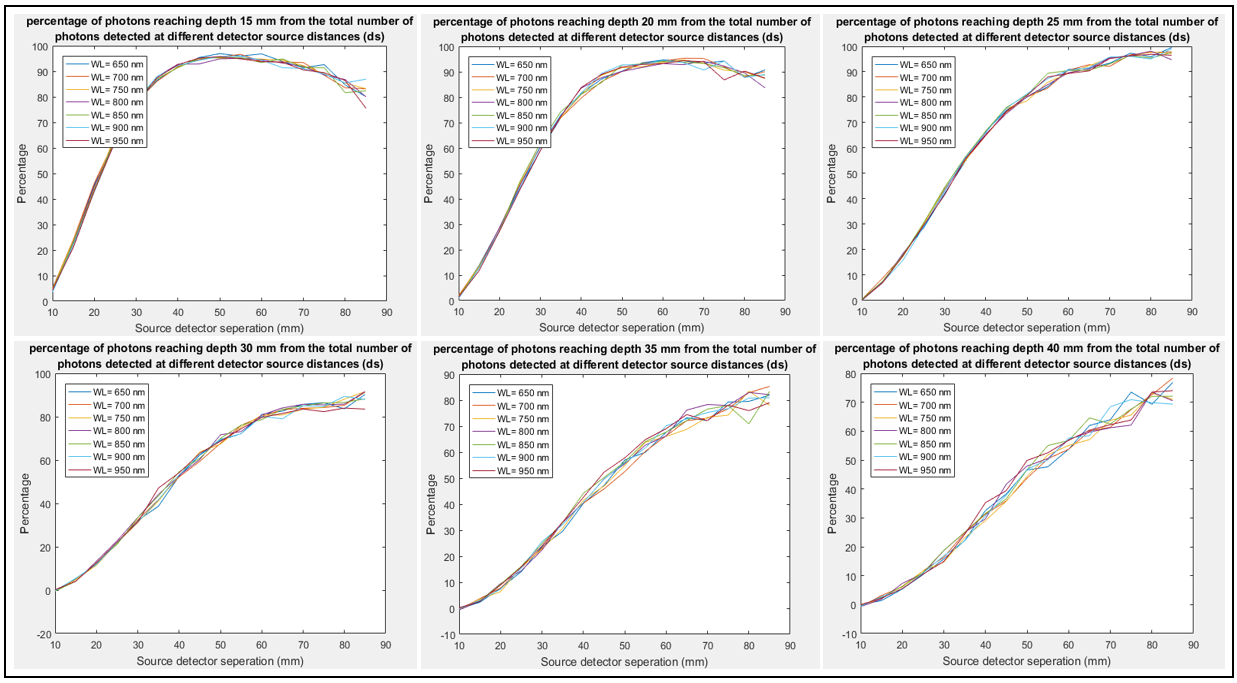}
	\caption{ The percentage of photons reaching depth of interest from the total number of photons detected at each detector
}
	\label{fig:photons_percent}
\end{figure*}

Figure \ref{fig:photons_reaching} and \ref{fig:photons_percent} illustrate the source detector distance and wavelength effect on penetration depth. Figure \ref{fig:photons_reaching} focuses on the ratio of photons reaching the depth of interest to the overall number of input photons. On the other hand, figure \ref{fig:photons_percent} provides a measure of sensitivity by presenting the percentage of detected photons which are reaching the depth of interest to overall detected photons at the same detector.

The comparison of figure \ref{fig:photons_reaching} and figure \ref{fig:photons_percent} offers a better understanding of optimal source-detector spacing. Although a higher number of photons can be detected at low spacings, a stronger signal to noise ratio is achievable using higher separation distances.

\begin{figure}\centering
	\includegraphics[width=8cm]{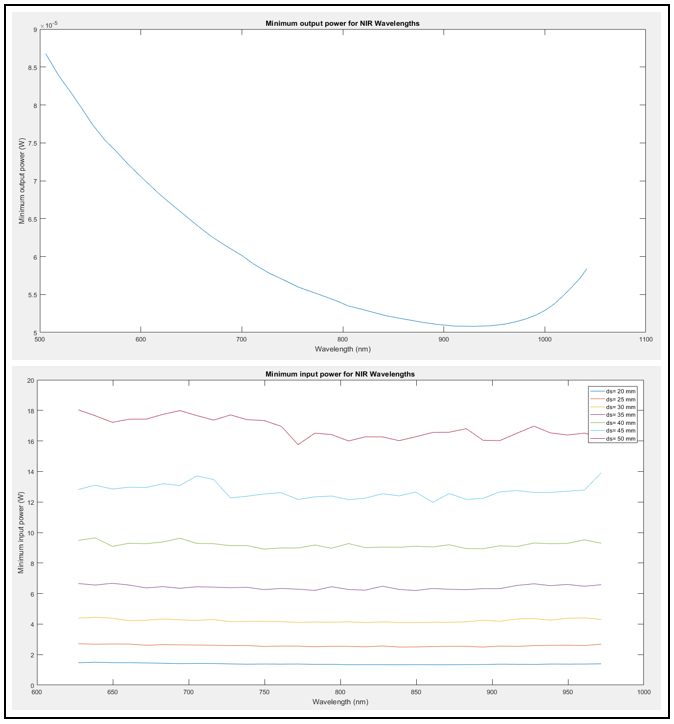}
	\caption{Minimum input power calculation using minimum sensible output power and power transmission ratio.}
	\label{fig:power}
\end{figure}

We present the minimum sensible output power and the resulted minimum input power in figure \ref{fig:power}.

\section{Conclusion}

In this work we performed DSE to analyze the effect of parameters such as wavelength and source-detector spacing on the penetration depth of NIR photons for a bladder volume spectroscopy application. We used monte carlo simulation using 7 different wavelengths and 16 possible source-detector spacings for 6 penetration depths. We have also illustrated the minimum input power for a specific choice of detector at these wavelengths and spacings. The results of our simulation can be used to direct the optimal design and implementation of a NIRS probe for bladder filling sensation.


\begin{thebibliography}{16}

\bibitem{2014:macnab}
Macnab, Andrew J. "The evolution of near infrared spectroscopy in urology." Biomedical Spectroscopy and Imaging 3.4 (2014): 311-344.

\bibitem{2005:macnab}
Macnab, A. J., R. E. Gagnon, and L. Stothers. "Clinical NIRS of the urinary bladder–A demonstration case report." Journal of Spectroscopy 19.4 (2005): 207-212.

\bibitem{2014:molavi}
Molavi, Behnam, et al. "Noninvasive optical monitoring of bladder filling to capacity using a wireless Near Infrared Spectroscopy device." IEEE transactions on biomedical circuits and systems 8.3 (2014): 325-333.

\bibitem{2016:fairchild}
FairChild. "QSB34GR / QSB34ZR / QSB34CGR / QSB34CZR
Surface-Mount Silicon Pin Photodiode." QSB34GR / QSB34ZR / QSB34CGR / QSB34CZR datasheet, Sep. 2016.

\bibitem{1998:Simpson}
Simpson, C. Rebecca, et al. "Near-infrared optical properties of ex vivo human skin and subcutaneous tissues measured using the Monte Carlo inversion technique." Physics in Medicine \& Biology 43.9 (1998): 2465.

\bibitem{2018:Urology}
Urology care foundation. "What is Urinary Incontinence?", http://www.urologyhealth.org/urologic-conditions/urinary-incontinence/printable-version. Accessed 23 June. 2018.

\end{thebibliography}

\end{document}